\documentclass[12pt]{iopart}

\usepackage{iopams}  
\usepackage{graphicx}
\usepackage{color}

\begin{document}

\title[Shock mode]{PIC simulations of stable surface waves on a subcritical fast magnetosonic shock front}

\author{M. E. Dieckmann}
\address{Dept. of Science and Technology (ITN), Link\"oping University, Campus Norrk\"oping, SE-60174 Norrk\"oping, Sweden}
\ead{mark.e.dieckmann@liu.se}

\author{C.~Huete}
\address{Univ Carlos III Madrid, Grp Mecan Fluidos, Leganes 28911, Spain}

\author{F.~Cobos}
\address{Univ Castilla La Mancha, ETSI Ind, Ciudad Real 13071, Spain}

\author{A.~Bret}
\address{Univ Castilla La Mancha, ETSI Ind, Ciudad Real 13071, Spain}

\author{D.~Folini}
\address{Univ Lyon, ENS de Lyon, Univ Lyon 1, CNRS, Centre de Recherche Astrophysique de Lyon UMR5574
F-69230, Saint-Genis-Laval, France}

\author{B.~Eliasson}
\address{Univ Strathclyde, SUPA, Glasgow G4 0NG, Scotland, UK}

\author{R.~Walder}
\address{Univ Lyon, ENS de Lyon, Univ Lyon 1, CNRS, Centre de Recherche Astrophysique de Lyon UMR5574
F-69230, Saint-Genis-Laval, France}

\vspace{10pt}
\begin{indented}
\item[]August 2022
\end{indented}

\begin{abstract}
We study with particle-in-cell (PIC) simulations the stability of fast magnetosonic shocks. They expand across a collisionless plasma and an orthogonal magnetic field that is aligned with one of the directions resolved by the 2D simulations. The shock speed is 1.6 times the fast magnetosonic speed when it enters a layer with a reduced density of mobile ions, which decreases the shock speed by up to 15\% in 1D simulations. In the 2D simulations, the density of mobile ions in the layer varies sinusoidally perpendicularly to the shock normal. We resolve one sine period. This variation only leads to small changes in the shock speed evidencing a restoring force that opposes a shock deformation. As the shock propagates through the layer, the ion density becomes increasingly spatially modulated along the shock front and the magnetic field bulges out where the mobile ion density is lowest. The perturbed shock eventually reaches a steady state. Once it leaves the layer, the perturbations of the ion density and magnetic field oscillate along its front at a frequency close to the lower-hybrid frequency; the shock is mediated by a standing wave composed of obliquely propagating lower-hybrid waves. We perform three 2D simulations with different box lengths along the shock front. The shock front oscillations are aperiodically damped in the smallest box with the fastest variation of the ion density, strongly damped in the intermediate one, and weakly damped in the largest box. The shock front oscillations perturb the magnetic field in a spatial interval that extends by several electron skin depths upstream and downstream of the shock front and could give rise to Whistler waves that propagate along the shock's magnetic field overshoot. Similar waves were observed in hybrid and PIC simulations and by the MMS satellite mission.
\end{abstract}

%
\vspace{2pc}

\submitto{\PS}
%
\maketitle
%
%

\section{Introduction}

Shocks in collisionless plasma, in which effects due to Coulomb collisions between charged particles are negligible compared to collective electromagnetic forces, have been studied in the laboratory~\cite{Romagnani2008,Kuramitsu2011, Ahmed2013,Schaeffer2017,Fazzini2022}, in the Solar system~\cite{Johlander2016}, and by means of numerical simulations~\cite{Winske1988,Lembege1992,Lowe2003,Chapman2005, Burgess2007,Yang2012,Clark2014,Dieckmann2014,Burgess2016,Marcowith2016,Umeda2017,Dieckmann2017,Gueroult2017,Dieckmann2018a,Kobzar2021}. Shocks are important structures for the dissipation of energy in collisionless plasma (See~\cite{Marcowith2016} for a recent review). We consider here perpendicular nonrelativistic fast magnetosonic (FMS) shocks, for which the magnetic field upstream of the shock is oriented orthogonally to its normal and is amplified by the shock crossing. In the reference frame of the shock, the plasma is slowed down, heated, and compressed by the shock crossing. 

Fast magnetosonic shocks are categorized according to how the shock speed in the upstream frame of reference compares to the speed of the magnetohydrodynamic FMS wave. If the shock speed is less than 2.7 times the FMS speed~\cite{Marshall1955}, the shock is subcritical and its electric cross-shock potential can slow down the plasma to a speed below the FMS speed. Another estimate~\cite{Edmiston1984} sets this number to a value below 2.7. If the shock is faster, the particle distributions around the shock become nonthermal~\cite{Gedalin2023}. Such distributions, which can involve for example particle beams or an anisotropic temperature, give rise to instabilities that modify or destroy the shock. Perpendicular subcritical shocks are stationary in time in their rest frame and they have a thin transition layer, which enables accurate measurements of their position. 

The Magnetospheric Multiscale Spacecraft (MMS) mission detected ripples on the surface of the Earth's bow shock~\cite{Johlander2016}. Surface waves on plasma boundaries~\cite{Cramer1995,Joarder2006,Lysak2008} can be stimulated by perturbations. Perturbations develop out of drift instabilities~\cite{Dieckmann2018a,Forslund1970,Davidson1977,Daughton2003} if plasma flows along a stationary boundary like a tangential discontinuity. Perturbations can also grow near shock surfaces. Shock-reflected ions can move far upstream of the shock and drive waves~\cite{Burgess2007,McClements1997} letting shocks propagate into an upstream medium with a spatially nonuniform density and magnetic field direction. Lowe and Burgess~\cite{Lowe2003} found boundary waves on the shock in their two-dimensional hybrid simulation, which used a kinetic approximation for ions and described electrons with an inertialess fluid. These waves were distributed over a wide wavenumber interval and followed the dispersion relation of the Alfv\'en wave~\cite{Gekelman2011}. Burgess and his coworkers extended the simulation domain to three dimensions and examined the interplay of the ripples with ion-driven instabilities and the therefrom resulting waves~\cite{Burgess2016}.

Hybrid codes approximate electrons by an inertialess fluid. Resolving their full dynamics is more expensive but it allows particle-in-cell (PIC) simulations to model high-frequency processes, where electron dynamics matters. Several PIC simulation studies have addressed the interplay of subcritical~\cite{Kobzar2021} and supercritical collisionless shocks~\cite{Lembege1992,Yang2012,Umeda2017} with the shock-modified upstream plasma. These studies covered a wide range of magnetic field strengths and orientations relative to the shock normal. Alfv\'enic shock ripples accelerated electrons and led to the growth of high-frequency waves like Whistlers. The plasma was not fully thermalized after its passage through the shock and relaxed through secondary instabilities. On ion gyro-scales, the shock was immersed in a transition layer where many waves and plasma structures were coupled across different spatial and temporal scales. 

The aforementioned simulations and the pioneering hybrid simulation in Ref.~\cite{Winske1988} suggest that stable, oscillatory shock surface modes exist in collisionless plasma that can be excited by perturbations of the upstream plasma and travel along the shock surface. In this context, the term stable means that once these surface waves have been excited, their amplitude remains constant or decreases.

The fundamental requirement for the existence of a stable surface wave is the presence of a restoring force, which serves to counteract the effects of perturbations. In the context of gas dynamics shocks, the mode is an acoustic one and the restoring force is provided by a combination of tangential velocity conservation and the unbalanced pressure field, leading to the formation of oscillating shock ripple patterns. For an ideal gas, the amplitude of these oscillations decreases over time proportional to $t^{-3/2}$, or $t^{-1/2}$ in the strong-shock limit, as demonstrated by the works of Roberts~\cite{Roberts1945}, Freeman~\cite{Freeman1955}, and Zaidel~\cite{Zaidel1960}. The stability limits, beyond which the shock perturbation may experience non-decaying behavior or even exponential growth, have been extensively studied and documented, departing from the pioneering works of D'yakov~\cite{DYakov1954} and Kontorovich ~\cite{Kontorovich1957} and continuing to high-energy-density conditions~\cite{Wetta2018,Huete2020}. The extension of these results to magnetohydrodynamics may not be trivial since additional magnetic restoring forces may appear. The first work on the stability limits of FMS shock in ideal conditions, characterized by an ideal gas equation of state and a perfectly conducting gas, was performed by Gardner~\cite{Gardner1964}. However, further research is necessary to fully understand the transient evolution of these perturbations. If FMS shock perturbations are damped and the damping rate is lower than their oscillation frequency, they have the potential to generate surface waves, as has been observed at the Earth's bow shock and in PIC and hybrid simulations.

In the previous hybrid- and PIC simulations, the upstream perturbations were driven primarily by the shock-reflected ion beam. In such a setting, the driver of the shock boundary oscillations cannot be separated from the shock boundary, which is characterized by an overshoot of the magnetic amplitude and density over its values downstream of the shock. This separation is necessary if we want to compare the oscillations of the shock boundary to shock modes in (magneto-)hydrodynamic models, which are usually taken to be monochromatic in wavenumber space. We can decouple the driver from the shock boundary by perturbing it once and letting it relax.

We study the evolution of a perturbation of the shock front with one- and two-dimensional particle-in-cell (PIC) simulations. We align the uniform magnetic field of the upstream plasma with the direction of the 2D simulation that is perpendicular to the shock normal. The magnetic field direction is unresolved in the 1D simulations. A thermal pressure gradient drives an initially planar subcritical FMS shock. We follow it through a spatially uniform magnetized ambient plasma with a ratio between the electron thermal pressure and the magnetic pressure $\approx 0.55$. The shock propagates across a perturbation layer with a limited extent along the shock propagation direction. The number density of mobile electrons in this layer equals that of the surrounding plasma. The positive charge density in the perturbation layer is subdivided into two components. The number density of mobile ions is constant along the shock propagation direction and is equal to or less than that of the surrounding ambient plasma. An immobile positive charge cloud cancels out the negative net charge.
In the perturbation layer of the 2D simulations, the number density of the mobile ions varies in the direction perpendicular to the shock normal. By selecting a sinusoidally varying perturbation, we isolate a single surface mode. We obtain the following results. 

Shocks in the 1D simulations propagate at a lower speed through the perturbation layer and regain their initial speed after they leave it. Based on this finding and the structure of the perturbation layer in the 2D simulations, the position of the shock front in the direction of the average shock normal should vary sinusoidally along the front and its amplitude should grow with time for as long as the shock moves through the perturbation layer. The amplitude does, however, saturate after an initial growth phase and remains constant after that. The density at the shock overshoot and the magnetic field direction also become functions of the position along the shock front. Once the shock leaves the perturbation interval and enters the spatially uniform upstream plasma, the spatial, density- and magnetic field perturbations perform oscillations around their equilibrium values. Their oscillation frequency is just below the lower-hybrid frequency and involve also the shock mode, which separates the upstream from the downstream plasma. The shock mode and the surface wave thus form an oblique FMS mode near its resonance frequency where it becomes quasi-electrostatic. 

Our paper is structured as follows. Section~2 summarizes the numerical scheme used by the PIC code, the initial conditions of our simulations, the FMS mode and its coupling to lower-hybrid waves, and results from 1D simulations. Section~3 presents results from 2D simulations and Section~4 discusses our findings and their implications.

\section{The simulation code and its initial conditions}

\subsection{The simulation code EPOCH}

PIC codes approximate each plasma species $i$ by resolving its velocity distribution function with a cloud of computational particles (CPs). Each CP $j$ of species $i$ is characterized by a charge $Q_j$ and mass $M_j$ with a value $Q_j/M_j$, which matches the charge-to-mass ratio $Q_i/M_i$ of a particle of species $i$. Every CP has a position $\mathbf{x}_j$ and velocity $\mathbf{v}_j$ and, hence, a spatially localized current density $\propto Q_j \mathbf{v}_j$. The summation of the current density contributions of all CPs yields the macroscopic current density $\mathbf{J}(\mathbf{x},t)$. The electric field $\mathbf{E}(\mathbf{x},t)$, magnetic field $\mathbf{B}(\mathbf{x},t)$ and current density $\mathbf{J}(\mathbf{x},t)$ are defined on a numerical grid and the latter updates $\mathbf{E}(\mathbf{x},t)$ and $\mathbf{B}(\mathbf{x},t)$ according to discretized forms of Amp\`ere's law 
\begin{equation}
\nabla \times \mathbf{B} - \frac{1}{c^2}\frac{\partial \mathbf{E}}{\partial t} = \mu_0 \mathbf{J},
\label{Ampere}
\end{equation}
where $\mu_0, c$ are the vacuum permeability and speed of light, and Faraday's law
\begin{equation}
\nabla \times \mathbf{E} + \frac{\partial \mathbf{B}}{\partial t} = 0.
\end{equation}
The EPOCH code fulfills $\nabla \cdot \mathbf{B}=0$ and Gauss' law $\nabla \cdot \mathbf{E}=\rho/\epsilon_0$ ($\rho,\epsilon_0$: charge density and vacuum electric permittivity) to round-off precision~\cite{Esirkepov2001}. Once the electromagnetic fields have been updated, they are interpolated to the position of each CP and update its momentum according to the relativistic Lorentz equation. All particle velocity components and, hence, all components of $\mathbf{E}$, $\mathbf{B}$, and $\mathbf{J}$ are updated also in simulations that resolve fewer than 3 spatial dimensions. This computational cycle is repeated for as many time steps $\Delta_t$ as necessary to cover the time scale of interest. More details of the numerical scheme of the EPOCH code are given elsewhere~\cite{Arber2015}.

\subsection{Initial conditions}

\begin{figure}
\includegraphics[width=0.8\columnwidth]{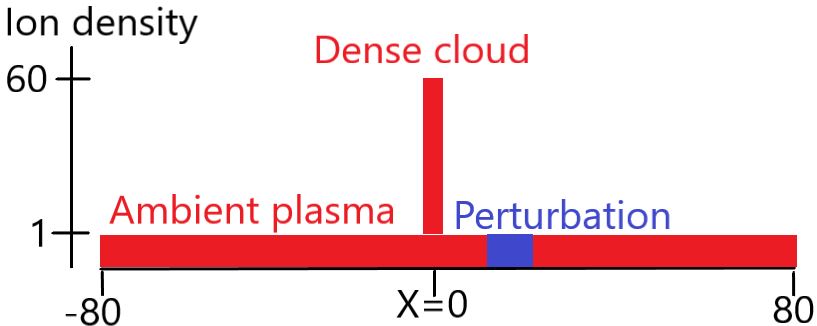}
\caption{The ion density distribution at the time $t=0$ in units of $n_{i0}$: The ambient plasma has the density 1. A dense cloud in the interval $-3 \le x \le 3$ has a density 60. The density of mobile ions in the perturbation layer $8.9 \le x \le 20.8$ is constant along $x$ but varies along $y$ in the 2D simulations. Its value varies between 0.4 and 1.0. In the 2D simulations, the initial ion density distribution does not change with $y$ outside of the perturbation layer.}
\label{figure01}
\end{figure}

All simulations cover the interval $-L_x/2 \le x \le L_x/2$. The 2D simulations also resolve an interval $0\le y \le L_y$. Values of $L_x$ and $L_y$ vary between simulations. Boundary conditions are periodic in all directions. We model fully ionized nitrogen at the correct ion-to-electron mass ratio because it is widely used in laser-plasma experiments, for example in Ref.~\cite{Ahmed2013}. Figure~\ref{figure01} sketches the initial number density distribution of the ions along $x$.
In the ambient plasma, we set the electron number density to $n_{e0}$ and that of the ions to $n_{i0}=n_{e0}/7$. The electron plasma frequency $\omega_{pe}={(n_{e0} e^2/\epsilon_0 m_e)}^{1/2}$ ($e, m_e, c$: elementary charge, electron mass, and light speed) sets the electron skin depth $\lambda_e = c/\omega_{pe}$, with which we normalize space. At the time $t=0$, the electric and magnetic fields are set to $\mathbf{E}=(0,~0,~0)$ and $\mathbf{B}=(0,B_0,0)$ with $eB_0/m_e \omega_{pe}=0.084$. All ions have a temperature 200~eV. Electrons outside (inside) the dense cloud have a temperature 1000~eV (1500~eV). Thermal diffusion lets more electrons stream from the dense into the ambient plasma than in the opposite direction, which yields an ambipolar electric field that points from the dense to the diluted plasma. This electric field lets the ions of the dense plasma expand into the ambient plasma. It also accelerates ambient electrons into the dense cloud. The accelerated ambient electrons form a beam that interacts with the electrons of the dense cloud. We give the latter a higher temperature in order to reduce the effects of a two-stream instability between both electron populations. 

The temperature $T_e =$ 1000 eV of the electrons in the ambient plasma sets their thermal speed  $v_{te}={(k_BT_e/m_e)}^{1/2}$ ($k_B$: Boltzmann constant). The ion temperature $T_i =$ 200 eV and $T_e$ set the ion acoustic speed $c_s$ in the ambient plasma. On the time scales of ion-acoustic oscillations, electrons have three degrees of freedom and ions have one. This gives the adiabatic constants $\gamma_e = 5/3$ for electrons and $\gamma_i = 3$ for ions. The ion-acoustic speed becomes $c_s ={(k_B(7 \gamma_e T_e + \gamma_i T_i)/m_i)}^{1/2}$. The Alfv\'en speed $v_A = B_0/(\mu_0 n_{i0} m_i)^{1/2}$ and $c_s$ define the FMS speed $v_{fms} ={(c_s^2+v_A^2)}^{1/2}$. Our plasma has the electron plasma beta $\beta = n_{e0}k_BT_e/(B_0^2/2\mu_0)=0.55$. Relevant plasma parameters in the uniform ambient plasma and their values are listed in Table~\ref{table1}. 
\begin{table}
\begin{tabular}{| l | l | l | } 
\hline
Electron gyrofrequency $\omega_{ce}=eB_0/m_e\omega_{pe}$ &  0.084 & $\omega_{ce}t_{sim}=1200$  \\
Ion gyrofrequency $\omega_{ci} = 7eB_0/m_i \omega_{pe}$ & $2.3\times 10^{-5}$ & $\omega_{ci}t_{sim}=0.33$   \\
Ion plasma frequency $\omega_{pi} = {(49e^2n_{i0}/\epsilon_0m_i)}^{1/2}/\omega_{pe}$ & 0.0165 & $\omega_{pi}t_{sim}=240$\\
Lower-hybrid frequency $\omega_{lh}={((\omega_{ce}\omega_{ci})^{-1}+\omega_{pi}^{-1})}^{-1/2}/\omega_{pe}$ & 0.0014 & $\omega_{lh}t_{sim}=20$  \\
\hline
Electron thermal speed $v_{te}={(k_BT_e/m_e)}^{1/2}/c$: & 0.044 & $v_{te}t_{sim}/\lambda_e=640$ \\
Ion-acoustic speed $c_s ={(k_B(7 \gamma_e T_e + \gamma_i T_i)/m_ic^2)}^{1/2}$ & $9.4\times 10^{-4}$ & $c_st_{sim}/\lambda_e=13.7$ \\
Alfv\' en speed $v_A = B_0/(\mu_0 c^2 n_{i0} m_i)^{1/2}$ & $1.4 \times 10^{-3}$ & $v_At_{sim}/\lambda_e =20.0$  \\
FMS speed $v_{fms} ={(c_s^2+v_A^2)}^{1/2}/c$: & $1.7\times 10^{-3}$ & $v_{fms}t_{sim}/\lambda_e=24.3$ \\
\hline
Electron thermal gyroradius $v_{te}/\omega_{ce}\lambda_e$ & $0.53$ & \\
Electron Debye length $v_{te}/\omega_{pe}\lambda_e$ & $0.044$ & \\
\hline
\end{tabular}
\caption{The relevant normalized frequencies, speeds, and spatial scales for the physical values $n_{e0}=10^{21}\mathrm{m^{-3}}$ and $B_0=0.85$ T we used in our simulation setup and for $t_{sim}=14500$ (in units of $\omega_{pe}^{-1}$).}
\label{table1}
\end{table}
Although we selected parameters in our simulation setup, which are representative of some laser-plasma experiments where physical scales matter, we use normalized units scaled to the electron skin depth $\lambda_e$, the electron plasma frequency $\omega_{pe}$, and the speed of light $c$.

We consider here FMS shocks that propagate through the ambient plasma. Wave dispersive properties of FMS shocks are set by the plasma conditions in the downstream plasma. However, the density and magnetic field amplitude do not vary much between the upstream (ambient) plasma and downstream plasma of a subcritical FMS shock. We discuss wave properties in the ambient plasma and assume that they are also representative of the downstream plasma. 

The FMS mode is not dispersive for low wavenumbers $k$. Since shocks form by a steepening of the wavefront, the wavenumbers of the waves that form the shock increase in time. Eventually, they reach values where the FMS mode becomes dispersive. Simplified expressions for the dispersion relation of FMS waves in collisionless plasma can be obtained by either neglecting space charge, which gives $\omega_{EM}(k) = v_{fms}k$, or by taking the electrostatic limit where the electric field is tied to oscillations of the charge density. The latter gives the lower-hybrid mode $\omega_{ES}(k) = (3v_{ti}^2k^2 + \omega_{pi}^2(\omega_{ce}^2+v_{te}^2k^2)/(\omega_{pe}^2+\omega_{ce}^2+v_{te}^2k^2))^{1/2}$ with the ion thermal speed $v_{ti}={(k_BT_i/m_i)}^{1/2}$. 

We use the noise distribution of the PIC code to track the full wave branch and show how it goes over into limits $\omega_{EM}(k)$ and $\omega_{ES}(k)$. Charge- and current-density fluctuations due to the moving CPs yield electric and magnetic field fluctuations. These fluctuations will have values of $k$ and $\omega$, which are connected to the particle motion. If CPs create fluctuations with such values of $k,\omega$, they can often also absorb them. Hence, strong fluctuations of the electric and magnetic fields tend to reveal locations in $k,\omega$-space where waves resonate with particles. They are strongest close to eigenmodes of the plasma (See~\cite{Dieckmann2004} for a related discussion). The FMS mode compresses the background magnetic field and we can use fluctuations of $B_y$ to track FMS modes in $k,\omega$-space. In our 1D geometry, fluctuations in $E_x$ are always tied to electrostatic charge density fluctuations and we use $E_x$ to identify the lower-hybrid mode in $k,\omega$-space.

\begin{figure}
\includegraphics[width=\columnwidth]{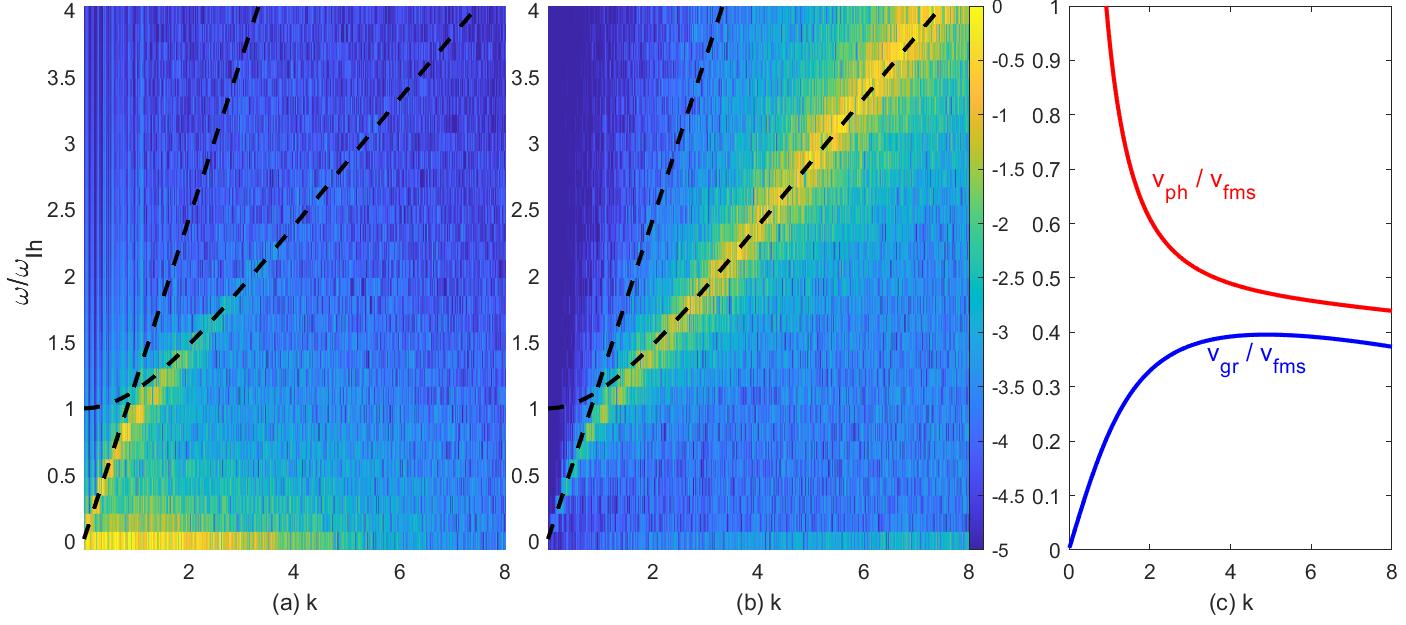}
\caption{Wave spectrum in the ambient plasma as a function of the wavenumber $k$ (normalized to $\lambda_e$) and the frequency $\omega$, which we normalize here to $\omega_{lh}$: Panel~(a) shows the power spectrum of the noise in the $B_y$ component and panel~(b) that of the $E_x$ component. The power spectra are normalized to their peak value and displayed on the same 10-logarithmic color scale. The dashed lines show the FMS mode $\omega_{EM}(k)$ and the lower-hybrid mode $\omega_{ES}(k)$. Panel~(c) plots the phase velocity $v_{ph}$ (red) and group velocity $v_{gr}$ (blue) computed from $\omega_{ES}(k)$.}\label{figure02}
\end{figure}

Figure~\ref{figure02} shows the power spectra of the fluctuations in the magnetic $B_y$ and electric $E_x$ components. We computed them by running a 1D PIC simulation that resolved $x$ with the plasma parameters of the ambient plasma discussed above. We Fourier-transformed $B_y(x,t)$ and $E_x(x,t)$ over space and time and multiplied the result with its complex conjugate giving $|B_y(k,\omega)|^2$ and $|E_x(k,\omega)|^2$.
The noise at low wavenumbers $k$ is magnetic and follows $\omega_{EM}$. At large $k$, the electric field noise maps out lower-hybrid waves. The waves that connect $\omega_{EM}$ with $\omega_{ES}$ at $k \approx 1$  have significant electric and magnetic components. They are thus not represented correctly by either limit.

Shocks form in our simulations just before the expanding ions reach the perturbation layer. We change the number density of mobile ions in this interval in Fig.~\ref{figure01} and keep that of mobile electrons unchanged. Since we set $\mathbf{E}= 0$ at $t=0$, Gauss' law $\nabla \cdot \mathbf{E}= \rho/\epsilon_0$ implies initially a zero total plasma charge density $\rho = 0$ everywhere, which places an immobile positive charge density in our perturbation layer. Its electric field cancels out the one caused by the jump in the number density of mobile ions.

The 1D simulations and the 2D simulations~1 and~2 resolve $L_x = 160$ with 8000 grid cells and end at the time $t_{sim}=14500$. The 2D simulation~3 uses 9000 grid cells to resolve $L_{x,3}=180$ and follows the shock for the time $t_{sim,3}=1.25t_{sim}$. The 2D simulations~1-3 resolve $L_{y,1}=12$ by 600 cells, $L_{y,2}=24$ by 1200 cells, and $L_{y,3} =36$ by 1800 cells. The data is averaged over 4 cells in the 1D simulations and over patches of $2\times 2$ cells in the 2D simulations. In the 2D (1D) simulations, ions and electrons are resolved by 25 CPs (375 CPs) per cell each.

\subsection{One-dimensional simulations}

We test with two 1D simulations how shocks react to changes in the number density of mobile ions in the perturbation layer. One 1D~simulation represents $0.7n_{i0}$ by mobile ions and the other 1D~simulation represents $0.4n_{i0}$ by mobile ions. According to Fig.~\ref{figure01}, one shock will be launched at the density jump to the right of the dense cloud and a second one at the density jump to its left. Both will propagate away from the dense cloud and into the ambient plasma. Only the shock that moves to the right propagates through the perturbation layer. The left-moving shock in one of the simulations is taken as the reference shock. We invert the sign of the position $x$ and velocity $v_x$ in our plots so that we can compare the ion phase space density distribution of the left-moving unperturbed shock with that of the right-moving perturbed shock.

\begin{figure}
\includegraphics[width=\columnwidth]{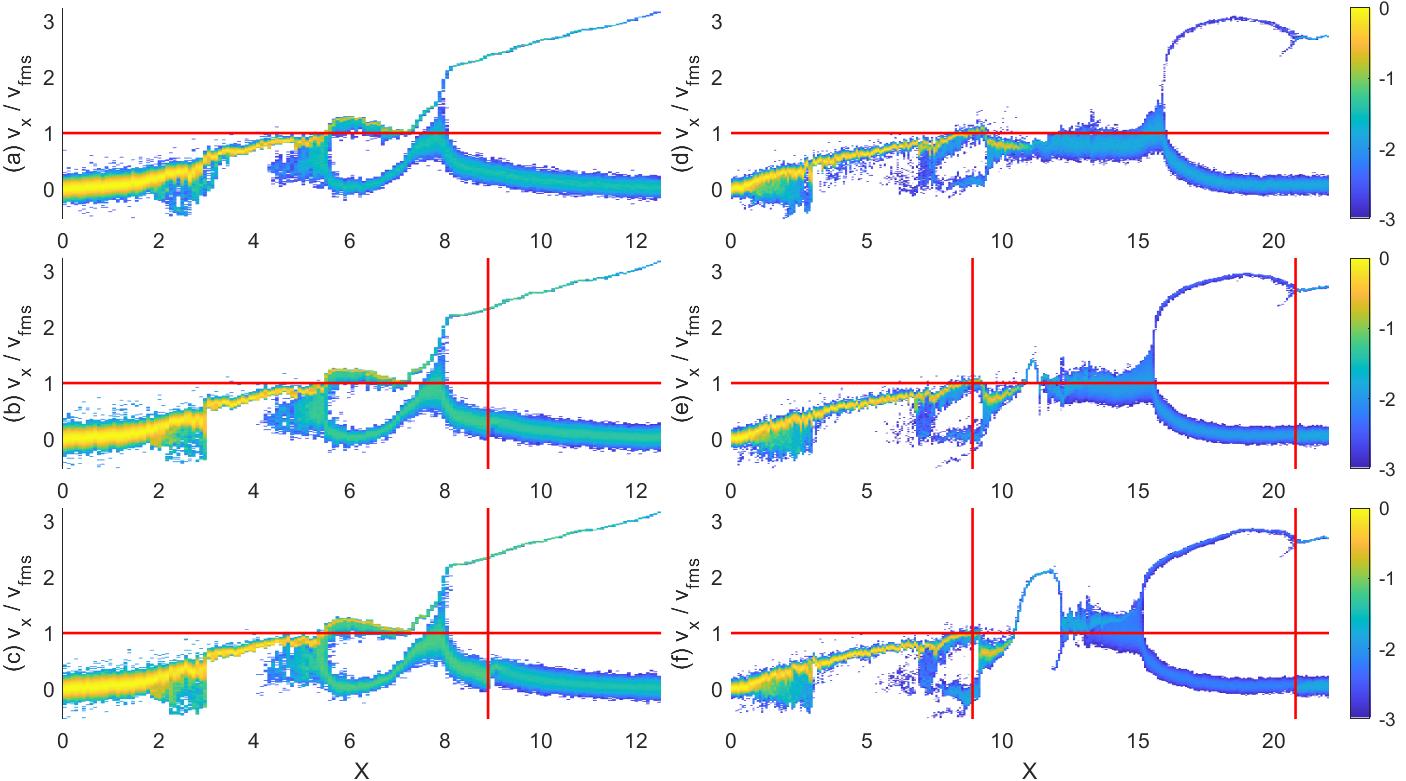}
\caption{Ion phase space density distributions at the time $t_1=2000$ (left column) and $t_2=5000$ (right column). Panels~(a,~d) correspond to the simulation with the unperturbed plasma. Panels~(b,~e) show the distributions for the simulation with a mobile ion density $0.7n_{i0}$ in the interval $8.9 \le x \le 20.8$. Panels~(c,~f) show the distributions for the simulation with a mobile ion density $0.4n_{i0}$ in the interval $8.9 \le x \le 20.8$. All distributions are normalized to the peak value at $t=0$ and shown on the same 10-logarithmic color scale. The vertical red lines indicate the boundaries of the perturbed interval while the horizontal ones plot the FMS speed.}\label{figure03}
\end{figure}

Figure~\ref{figure03} shows ion phase space density distributions. The supplementary movie~1 animates their evolution in time until $t_{sim}=14500$.
At $t_1=2000$, a localized oscillation of the ion phase space density distribution is growing at $x\approx 8$. We may interpret it as a steepening wave. It breaks shortly after this time and changes into a shock~\cite{Gueroult2017}. The magnetic field of the shock traps the electrons ahead of it and pushes them upstream. The current of the moving electrons and their space charge induces an electric field. It accelerates the practically unmagnetized ions with $\omega_{ci}t_{sim} \ll 2\pi$ ahead of the shock until their current balances the electronic one, giving rise to the shock foot. The change in the ion number density near the boundary $x=8.9$ of the perturbation layer in Figs.~\ref{figure03}(b,~c) is compensated by a faster motion of the ions in the perturbed interval. 

At $t_2=5000$ or $\omega_{lh}t_2=2.2\pi$, qualitative differences can be observed between the three shocks. Most downstream ions in the interval $12 \le x \le 16$ in Fig.~\ref{figure03}(d) move at a speed below $v_{fms}$ while the mean speed of those in Fig.~\ref{figure03}(e) is about $v_{fms}$. The downstream ions in Fig.~\ref{figure03}(f) are confined to a smaller interval $13 \le x \le 15$ and most ions move faster than $v_{fms}$. Despite its faster downstream ions, the shock at $x=15$ in Fig.~\ref{figure03}(f) is trailing those in Figs.~\ref{figure03}(d,~e). Fewer ions, which cross the shock, reduce the thermal pressure behind it and, hence, the shock speed in the downstream frame. Figure~\ref{figure03}(f) also shows that the slowdown of the perturbed shock decreased the speed of the reflected ions. The ions at $x\approx 20$, which were reflected by the shock when it just entered the perturbation layer, have a speed $\approx 3v_{fms}$. Those at $x\approx 17$, which were reflected after the slowdown of the shock, reach a speed of only $2.2v_{fms}$. 

In Fig.~\ref{figure03}(d), the mean velocity of the ions is approximately constant for $9.5 \le x\le 16$. The ion density change at $x\approx 11$ is caused by a tangential discontinuity, which balances the high thermal pressure of the blast shell plasma against the combined thermal and magnetic pressure of the shocked ambient plasma. As the shocks in Figs.~\ref{figure03}(e,~f) move into the perturbed interval, the pressure ahead of the tangential discontinuity decreases. Blast shell ions accelerate for $10 \le x \le 11$ in Figs.~\ref{figure03}(e,~f) in the ambipolar electric field of the blast shell's density gradient and decelerate at larger $x$, transferring their momentum to the downstream plasma.

Figures~\ref{figure03}(d,~e,~f) show a structure in the shock-reflected ion beam at $x\approx 21$ and $v_x/v_{fms}\approx 2.8$. It develops when the shock-reflected ion beam, which has not yet developed at the snapshots for $t_1=2000$, catches up with the ions at the front of the rarefaction wave that is visible in Figs.~\ref{figure03}(a,~b,~c) at high speeds for $x>8$ (See supplementary movie~1).

Figure~\ref{figure04} shows the ion phase space density distributions at later times. We also plot the distributions of the ion densities that correspond to those of the phase space density.
\begin{figure}
\includegraphics[width=\columnwidth]{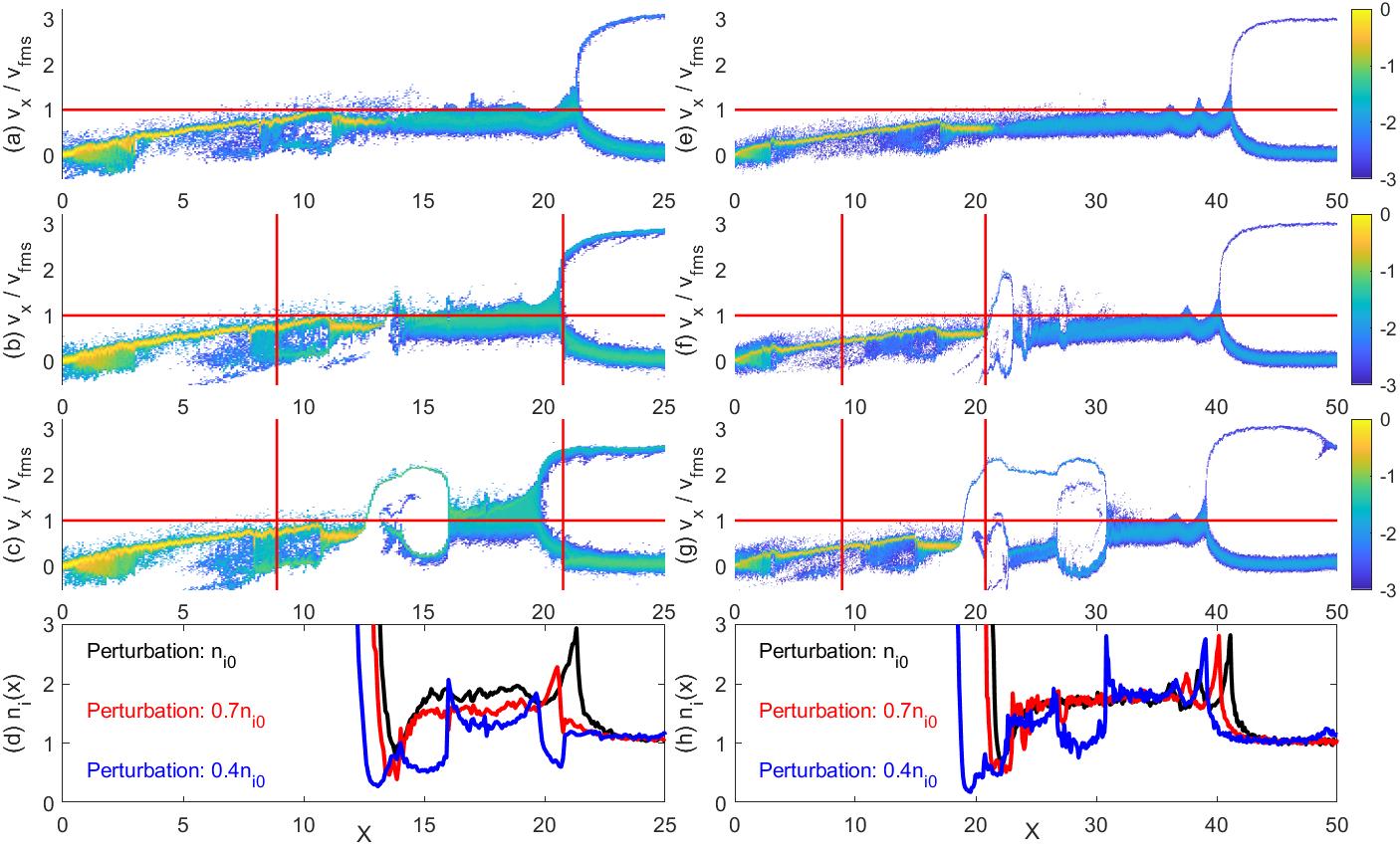}
\caption{Ion density and phase space density distributions at the time $t_3=7000$ (left column) and $t_{sim}=14500$ (right column). Panels~(a,~e) correspond to the simulation with the unperturbed plasma. Panels~(b,~f) show the distributions for the simulation with a mobile ion density $0.7n_{i0}$ in the interval $8.9 \le x \le 20.8$. Panels~(c,~g) show the distributions for the simulation with a mobile ion density $0.4n_{i0}$ in the interval $8.9 \le x \le 20.8$. All distributions are normalized to the peak value at $t=0$ and shown on the same 10-logarithmic color scale. Panels~(d,~h) plot the densities of the mobile ions at the times $t_3$ and $t_{sim}$, respectively.}
\label{figure04}
\end{figure}
At $t_3=7000$ ($\omega_{lh}t_3 =3.1\pi$), the shocks are about to leave the perturbation layer. As before, the mean velocity of the downstream ions increases, and the shock speed in their rest frame decreases with a decreasing number density of the mobile ions ahead of the shock. A substantial lag is observed in particular for the shock in Fig.~\ref{figure04}(c). The ion densities for this time are plotted in Fig.~\ref{figure04}(d). In the overshoot of the unperturbed shock, the ion density reaches the value 3. It decreases to just below 2 downstream of the shock. The shocks, which move through the perturbation layer, have a lower density of their downstream plasma, and the ion density at their overshoot reaches about 3 times that of the mobile ions in the perturbation layer. At $t_{sim}=14500$ ($\omega_{lh}t_{sim} =6.4\pi$), the shock speed and the mean velocity of the downstream ions behind the shocks are similar. The ion phase space vortices~\cite{Eliasson2006} in Fig.~\ref{figure04}(f,~g) have been separated from the shock by the influx of shocked upstream plasma. The density distributions near the overshoots in Fig.~\ref{figure04}(h) are similar for all shocks apart from a displacement along $x$.

\begin{figure}
\includegraphics[width=\columnwidth]{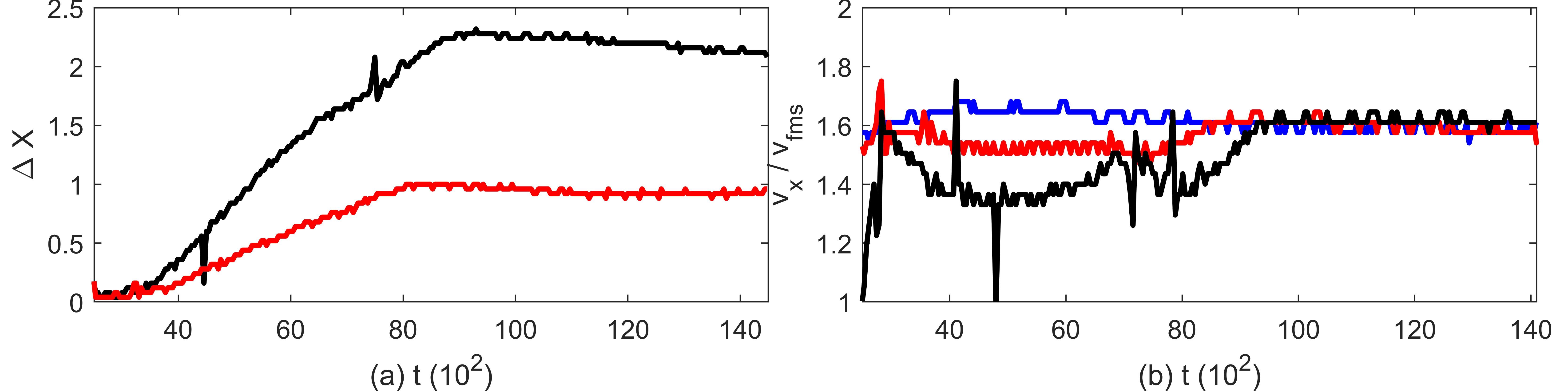}
\caption{Time evolution of the shock separation and speed: Panel~(a) plots the separation of the unperturbed shock and the one that propagated through the ion perturbation with the density $0.7n_{i0}$ (red curve). The black curve shows the lag of the shock that propagated through the ion perturbation with the density $0.4n_{i0}$. Panel~(b) plots the speed of the shock that propagates through the unperturbed plasma (blue), through the one with $0.7n_{i0}$ mobile ions (red), and with $0.4n_{i0}$ (black).}
\label{figure05}
\end{figure}

Figure~\ref{figure05} quantifies the impact of the changed density of mobile ions on the shock speed and position. We identify the position, where the shock is located, as the one with the largest curvature of the ion density and track it over time. We averaged the ion density over several grid cells to decrease statistical fluctuations of the ion density, which reduces the accuracy with which we can determine the spatial position of the shock's overshoot. The method, with which we determined the shock position, does also not always find the exact position of the shock in particular in the perturbation layer with the density of mobile ions $0.4n_{io}$, which created several outliers in the data. Inaccuracies will be visible in particular in the velocity data that is obtained by differentiating the noisy position data. Nevertheless, the computed curves reveal trends that are confirmed by the supplementary movie~1. Figure~\ref{figure05}(a) plots the separation of the reference shock, which moves through the unperturbed plasma, from the two shocks that move through the perturbed plasma. We start plotting the separation after $t=2500$ when the shock fully reformed in Figs.~\ref{figure03}(a-c).
The shock moving through the perturbation layer with the mobile ion density $0.7n_{i0}$ falls steadily behind the unperturbed shock until $t\approx 8000$ when Fig.~\ref{figure04}(b) shows that it crosses the upper boundary $x=20.8$ of the perturbed interval. The shock, which moves through the plasma with the mobile ion density $0.4n_{i0}$, leaves the perturbation layer last. Hence it is slowed down until $t\approx 10^4$. Figure~\ref{figure05}(b) obtains the shock speed from the change of its position over a time interval $\approx 640$. The speeds of the shocks remain above $1.3 v_{fms}$ for all times. Given that the FMS wave with frequencies just below $\omega_{lh}$ is dispersive and that its phase speed is well below $v_{fms}$ at wave numbers $k\approx 1$ in  Fig.~\ref{figure02}, the Mach numbers are higher.

\section{Two-dimensional simulations}

\subsection{A comparison of the boundary oscillations in the 2D simulations}

In what follows, the term "simulation $j$" with $1 \le j \le 3$ refers to the 2D simulation with the length $L_{y,j}$ along $y$. We express the number density of mobile ions $n_{i}(x,y)$ in units of $n_{i0}$ and the in-plane magnetic field is $B(x,y)={(B_x^2+B_y^2)}^{1/2}/B_0$. The magnetic $B_z$ component remains at noise levels. The perturbation layer covers again $8.9 \le x \le 20.8$. Within the perturbation layer of simulation~$j$, the density of mobile ions varies as $n_i(x,y) = 0.7+0.3\sin{(2\pi y / L_{y,j})}$ for $0 \le y \le L_{y,j}$. We take the shock that moves in the direction of decreasing $x<0$ as the unperturbed reference shock.

\begin{figure}
\includegraphics[width=\columnwidth]{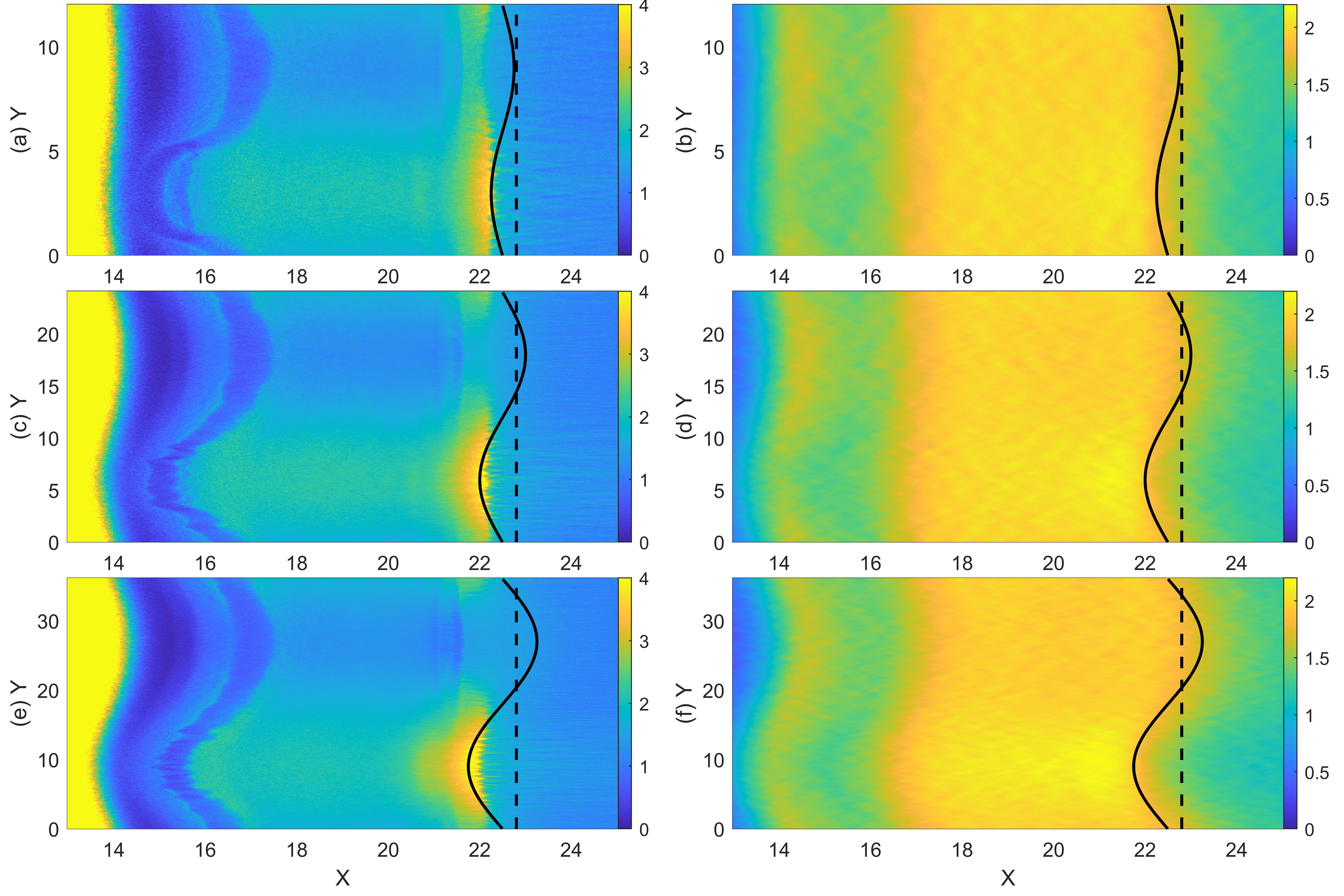}
\caption{Ion density $n_i(x,y)$ and in-plane magnetic field amplitude $B(x,y)$ at $t=7000$: Panels (a)~and~(b) show $n_i(x,y)$ and $B(x,y)$ computed by simulation~1 and panels~(c)~and~(d) those of simulation~2. Panels~(e)~and~(f) show the $n_i(x,y)$ and $B(x,y)$ computed by simulation~3. The dashed line denotes the average position of the unperturbed shock and the solid line corresponds to $x_c(y)$.}
\label{figure06}
\end{figure}

Figure~\ref{figure06} shows $n_i(x,y)$ and $B(x,y)$ at the time $t=7000$, when the perturbed shocks in the three simulations have left the perturbation layer and entered the spatially uniform ambient plasma. Figures~\ref{figure06}(a,~c,~e) show in each simulation $j$ a localized peak of the ion density at $x\approx 22$. All shocks lag behind the reference shock. The shape of the density peak varies between the simulations. With respect to the upstream region, the shock front is concave near $y=10$ in simulation~3 and follows the isocontour of the magnetic amplitude. It is almost planar at $y\approx 5$ in simulation~2 and convex at $y\approx 3$ in simulation~1. We also observe a structure with a reduced ion density, which is a signature of an ion phase space vortex like the ones shown in Fig.~\ref{figure04}, at $x\approx 16$ and a non-planar front of the dense blast shell for $x\approx 14$. 

Figures~\ref{figure06}(b,~d,~f) show that the distributions of $B(x,y)$ react to the density modulations of the shock fronts. The varying density of mobile ions in the perturbation layer affected the balance between the thermal- and magnetic pressures downstream of the shock and the ram pressure of the upstream ions. The magnetic field expanded in the upstream direction in intervals with a low density of mobile ions. We observe bent magnetic field lines at the shock and around the ion phase space vortex. The curves $x_{c,j}(y)=22.5-A_j\sin{(2\pi y/L_{y,j})}$ with $A_j/L_{y,j}= 0.021$ follow the modulation of the front of $B(x,y)$. 
The supplementary movie~2 (simulation~1), movie~3 (simulation~2), and movie~4 (simulation~3) show the evolution of the shocks in the perturbation layer. It demonstrates that the density and magnetic field perturbations grow, saturate, and remain stationary until they leave it. This differs from what we observed in Fig.~\ref{figure05}(a) where the distance between the reference shock and the perturbed shock steadily increased until the shock left the perturbation layer.

The density oscillation along $y$ and around $x\approx 22$ in Fig.~\ref{figure06}(a,~c,~e) induces an electric field that points from the dense to the dilute plasma and accelerates ions. It acts as a standing surface wave with a wavenumber $k_{y,j}=2\pi/L_{y,j}$. Electrons can move freely along the magnetic field and the perturbation could thus oscillate in the ion-acoustic mode with the frequency $\omega_{cs,j} = c_s k_{y,j}$. If it does, it will oscillate with $\omega_{cs,1} \approx 0.36\omega_{lh}$, $\omega_{cs,2}=0.24\omega_{lh}$, and $\omega_{cs,3}=0.12\omega_{lh}$ in simulations~1 to~3. The density perturbation along $y$ is also coupled to the shock with its average normal along $x$. The upstream plasma that crosses the shock is compressed by a FMS mode with a frequency close to $\omega_{lh}$. The large difference between the ion-acoustic frequency and $\omega_{lh}$ allows us to distinguish between both.

\begin{figure}
\includegraphics[width=\columnwidth]{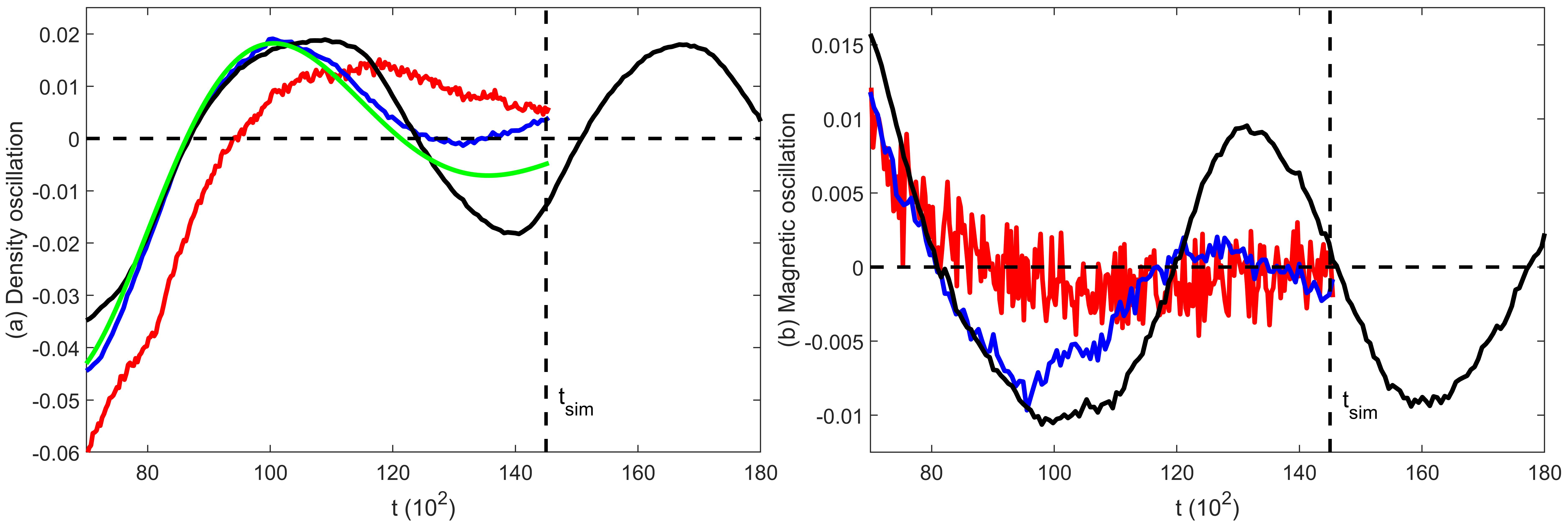}
\caption{Oscillations of the ion peak density $n_{i,max}(k_{y,j},t)$ are plotted in panel~(a) and the isocurves $x_B(k_{y,j},t)=1.67$ of $B(x,y)$ are plotted in panel~(b). The red curves correspond to simulation~1, the blue ones to simulation~2, and the black curves to simulation~3. The curve $C(t)=-0.044\cdot \cos{(\omega_ft-0.2\pi)}\cdot \exp{(-0.3\omega_ft)}$ with $\omega_f =0.76\omega_{lh}$ is plotted in green in panel~(a).}
\label{figure07}
\end{figure}

For every data time step, we determine the maximum value of the ion density along $x$ as a function of $y$, which gives us a density distribution $n_{i,max}(y,t)$. We also determine for each $y$ the location where we reach $B(x,y)=1.67$ first coming from the upstream region, which gives us the evolution in time of a curve $x_B(y,t)$ that is similar to the solid curve in Figs.~\ref{figure06}(b,~d,~f). The imaginary part of the fundamental frequency of the Fourier transform over space of $x_B(y,t)$ and $n_{i,max}(y,t)$ gives us the amplitudes $x_B(k_{y,j},t)$ and $n_{i, max}(k_{y,j},t)$. Figure~\ref{figure07} plots $x_B(k_{y,j},t)/L_{y,j}$ and $n_{i,max}(k_{y,j},t)/L_{y,j}$.
In simulation~1 $x_B(k_{y,j},t)$ and $n_{i,max}(k_{y,j},t)$ are strongly damped. Both curves reach extrema at around $t=1.1\times 10^4$ and converge to a steady state after that. Weaker damping in simulation~2 yields pronounced extrema at $t\approx 10^4$ and weaker ones at $t\approx 1.3 \times 10^4$. Both curves in simulation~3 oscillate in antiphase with the period $\approx 6000$. The oscillation frequency is lowest in simulation~1 and it is similar in simulations~2 and~3, which is not what we expect from ion-acoustic oscillations with $\omega_{cs,1}=2\omega_{cs,2}$ and $\omega_{cs,1}=3\omega_{cs,3}$. Simulation~3 gives us the oscillation frequency $0.75\omega_{lh}$ or $6.25\omega_{cs,3}$. 

Lower-hybrid waves have been analyzed in uniformly magnetized plasma~\cite{Verdon2009a,Verdon2009b}. They trap electrons magnetically and the trapped electrons confine the ions electrically. Lower-hybrid waves require that the trapped electrons do not move far along the magnetic field during $2\pi\omega_{lh}^{-1}$. Their wavevector, which is aligned with the density gradient, therefore has to be almost perpendicular to the magnetic field. The perturbation in our 2D simulation rotates the magnetic field direction and the density gradient differently and both will not remain mutually orthogonal. This rotation sets an upper limit on the amplitude of the spatial oscillations of the magnetic field and, hence, on the perturbation amplitude in our 2D simulations. 

The angle $\theta$ between the wavevector of an undamped lower-hybrid mode and the magnetic field is limited to the range $|90^\circ - \theta | \lesssim \theta_{max}$. In a plasma with equal temperatures of electrons and ions, the angular range can be estimated with $\cos^2\theta_{max}=m_e/m_i$, which yields $\theta_{max}\approx 0.4^\circ$ for nitrogen ions. 

The perturbed shock propagates into an upstream plasma, which is spatially homogenous and does therefore not favor a specific propagation direction of the boundary oscillation. Hence, the surface wave and the mode that compresses the upstream plasma constitute a standing wave, which is composed of modes that propagate obliquely to the shock normal and in opposite directions. We can estimate the propagation angle $\theta_j$ of these modes relative to the background magnetic field as follows. According to the ion velocity distribution in the interval $37 \le x \le 41$ in Fig.~\ref{figure04}({\color{blue}e}), the wavelength of the lower-hybrid mode, which sustains the shock and forms the component of the wave that points along its normal, is $k_{x,s}\approx \pi$. The wavelength of the surface wave that moves along the shock boundary is $k_{y,j}=2\pi/L_{y,j}$. The propagation angles $\theta_j = k_{x,s}/k_{y,j}$ are thus $\theta_1 = 80.5^\circ$, $\theta_2 = 85.2^\circ$, and $\theta_3 = 86.8^\circ$. 

Shock modes do not have to be undamped as they are continuously fed with energy from the inflowing upstream plasma. The wave vector of the shock mode may however be rotated into a direction with less damping. Another aspect is that lower-hybrid waves with frequencies $\omega \approx \omega_{lh}$ and with $\theta = 90^\circ$ are dispersive~\cite{Dieckmann2017}. If their phase velocity also changes with $\theta$, we obtain an undamped surface wave only if the spread in wavenumbers along $k_x$ and $k_{y}$ is small. The higher resolution of $k_y$ by simulation~3 allows the surface wave to involve wave modes with similar frequencies. Given that the shock fronts involve more than one wave mode, different spectral resolutions may also explain why their shapes differ in Figs.~\ref{figure06}(a,~c,~e).

In addition to the requirement that lower-hybrid waves can only mediate a shock if they propagate almost perpendicularly to the background magnetic field, magnetic tension $(\mathbf{B}\cdot \nabla ) \mathbf{B}/\mu_0$ is likely to contribute to the saturation of the perturbation. In simulation~3 and at the time depicted in Fig.~\ref{figure06}(f), its magnitude is comparable to that of the magnetic pressure gradient force density $\nabla \mathbf{B}^2/2\mu_0$ at $x\approx 22$ and $y\approx 9$ and about 20\% of the thermal pressure gradient force density at the front of the shock in Fig.~\ref{figure06}(e) at the same position (not shown). As we will see, the magnetic tension builds up over a much wider x-interval than the other forces and its impact on the saturation of the shock deformation in the perturbation layer is thus difficult to quantify.

\subsection{Damping of the shock oscillations in simulation 1 and 2}

\begin{figure}
\includegraphics[width=\columnwidth]{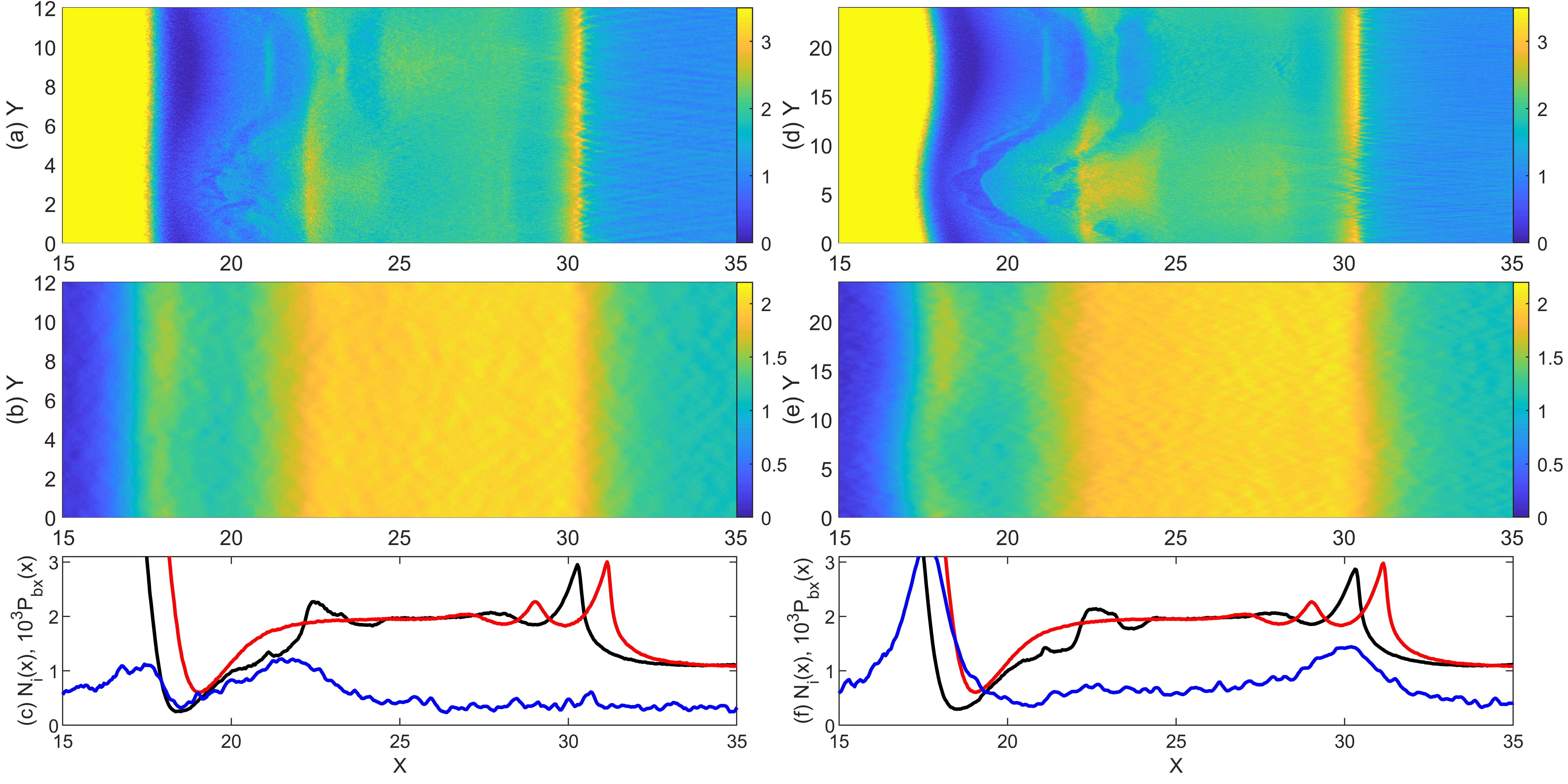}
\caption{Distributions of the ion density $n_i(x,y)$ and the amplitude of the in-plane magnetic field $B(x,y)$ in simulations 1 (left column) and 2 (right column) at the time $t=10^4$: Panels (a) and (d) show the ion density distributions. Panels (b) and (e) show $B(x,y)$. Panel~(c) and~(f) plot the box-averaged $N_i(x)=L_y^{-1}\int n_i(x,y) \, dy$ of the perturbed shock (black curve), of the reference shock (red curve), and the box-averaged magnetic pressure component $P_{bx} = L_y^{-1} \int (B_x(x,y)/B_0)^2 \, dy$ of the perturbed shock multiplied with $10^3$ (blue curve).}
\label{figure08}
\end{figure}

We examine the distributions of the ion density and magnetic field amplitude at two times. Figure~\ref{figure08} shows them for simulations 1 and 2 at the time $t=10^4$, when the curves for simulation~2 went through their extrema in Fig.~\ref{figure07}.
The perturbation of the shock front in simulation~1, which is shown in Figs.~\ref{figure08}(a, b),  has damped out. The front is planar and the density of the overshoot and the magnetic amplitude does not vary with $y$. The front is located one electron skin depth behind that of the reference shock and $P_{bx}$, which quantifies the deformation of the magnetic field in the simulation plane, is at noise levels. The perturbation of the shock front in simulation~2 has the opposite phase than the initial one in Fig.~\ref{figure06}(c,~d); it is oscillating. In simulation~2, $P_{bx}$ evidences a deformation of the magnetic field, which results in magnetic tension, in an interval with the width 3 centered on the shock. 

\begin{figure}
\includegraphics[width=\columnwidth]{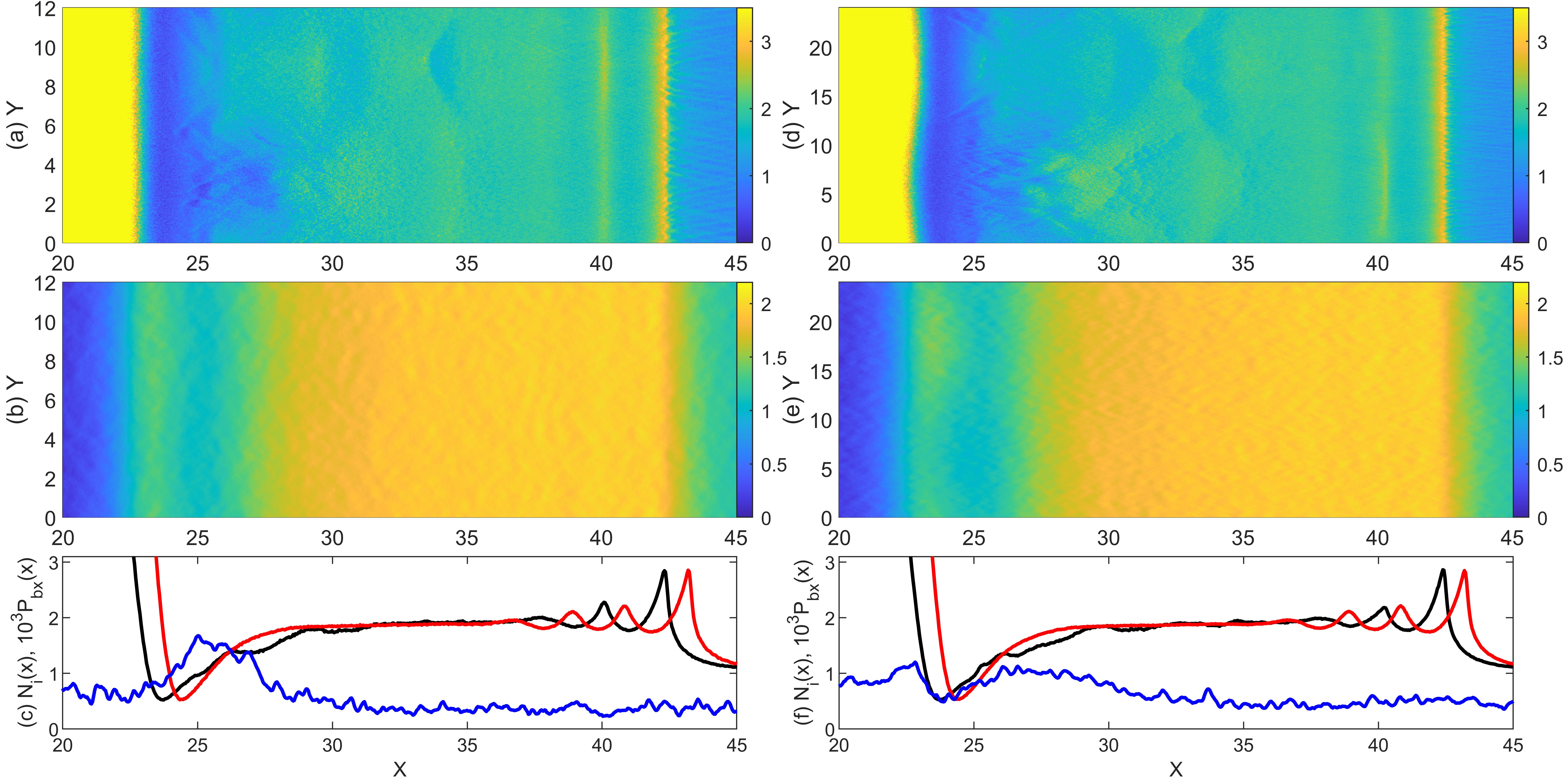}
\caption{Distributions of the ion density $n_i(x,y)$ and the amplitude of the in-plane magnetic field $B(x,y)$ in simulations 1 (left column) and 2 (right column) at the time $t_{sim}$: Panels (a) and (d) show the ion density distributions $n_i(x,y)$. Panels (b) and (e) show $B(x,y)$. Panel~(c) and~(f) plot the box-averaged $N_i(x)=L_y^{-1}\int n_i(x,y) \, dy$ of the perturbed shock (black curve), of the reference shock (red curve), and the box-averaged magnetic pressure component $P_{bx} = L_y^{-1} \int (B_x(x,y)/B_0)^2 \, dy$ of the perturbed shock multiplied with $10^3$ (blue curve).}
\label{figure09}
\end{figure}

Figure~\ref{figure09} shows that at $t=t_{sim}$, the shock perturbation has also damped out in simulation~2. The density distribution of simulation~1 in Fig.~\ref{figure09}(a) shows another density maximum at $x\approx 40$. Its separation from the leading density maximum at $x\approx 42$ reveals that the wavelength of the lower-hybrid mode, which mediates the shock, is 2 and $k_{x,s}=\pi$ as we observed in the 1D simulations. 
The magnetic field near the shocks in both simulations is aligned with $y$ at this time. The lag between the perturbed and unperturbed shocks in Figs.~\ref{figure09}(c,~f) is unchanged compared to that in Figs.~\ref{figure08}(c,~f). 

\subsection{Weakly damped oscillations in simulation 3}

\begin{figure}
\includegraphics[width=\columnwidth]{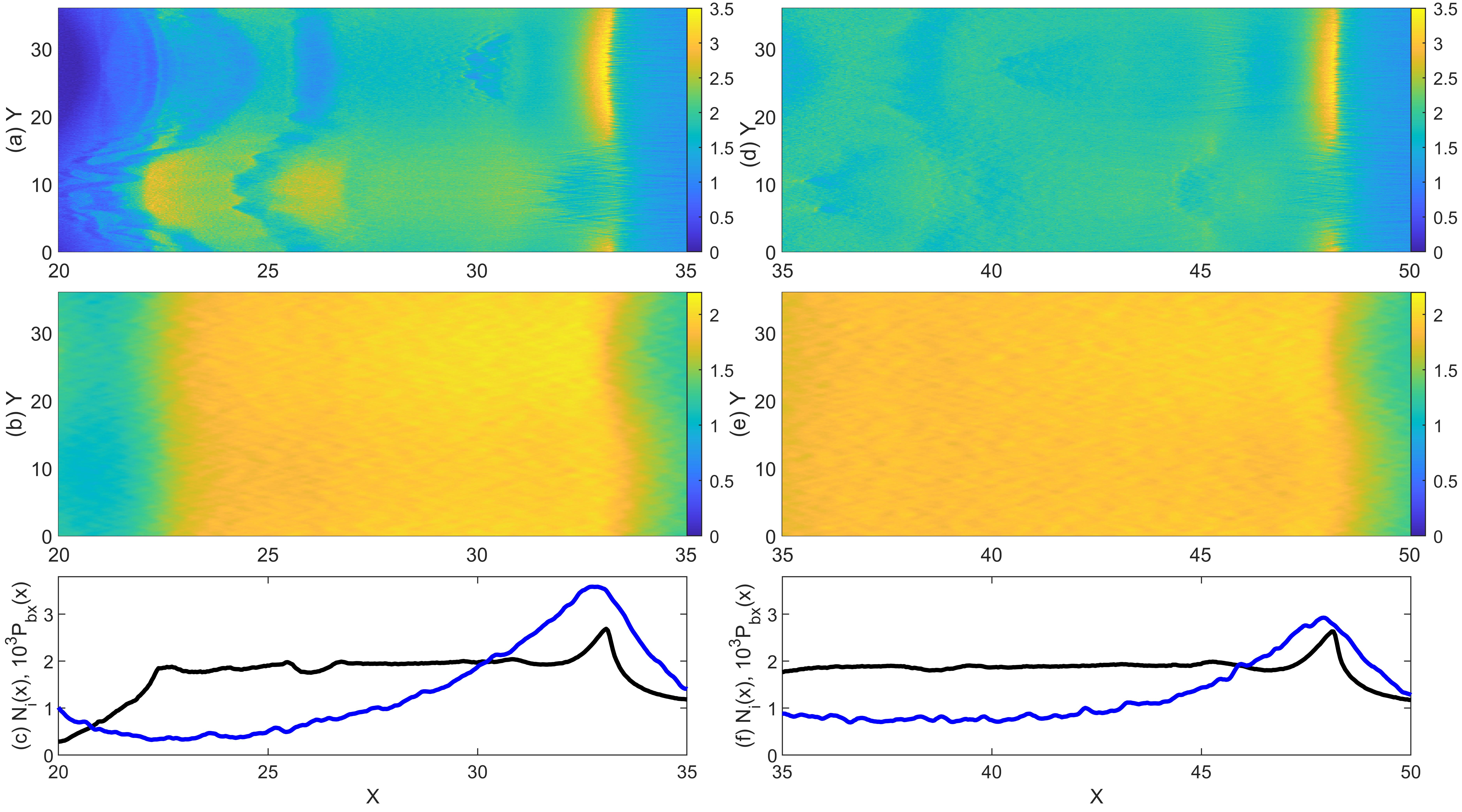}
\caption{The plasma states at the times $t=1.1 \times 10^4$ (left column) and $1.66 \times 10^4$ (right column) in simulation~3: Panels (a,~d) show the ion density distributions $n_i(x,y)$. Panels~(b,~e) show the amplitude of the in-plane magnetic field $B(x,y)$. Panels~(c) and~(f) show the y-averaged ion density $N_i(x)=L_y^{-1}\int n_i(x,y) \, dy$ (black curves) and $P_{bx} = L_y^{-1} \int (B_x(x,y)/B_0)^2 \, dy$ (blue curves).}
\label{figure10}
\end{figure}

Figure~\ref{figure10} shows the ion density and magnetic field distributions at the times $t=1.1 \times 10^4$ and $1.66 \times 10^4$, when the curves $n_i(k_{y,3},t)$ and $x_B(k_{y,3},t)$ go through extrema in Fig.~\ref{figure07}. 
The oscillations of the shock fronts are in phase at the times $t=1.1 \times 10^4$ and $t=1.66 \times 10^4$ in simulation~3. Figure~\ref{figure10}(a) reveals a density maximum at $x\approx 33$ and $y>18$, which is concave with respect to the upstream region. A weak density enhancement near $y\approx 8$ bulges out into the upstream region. The position of the shock front is thus also a function of $y$ and the variation is sinusoidal. The density oscillation is still present in Fig.~\ref{figure10}(d). The front is almost flat, which indicates that some of the waves that sustained the shock have subsided. The in-plane magnetic field $B(x,y)$ in Figs.~\ref{figure10}(b,~e) is deformed near the shock front. It expands upstream in intervals, where the shock density is low. The supplementary movie~4 confirms that the shock front performs sinusoidal oscillations around its mean position along $x$, which are synchronized with those in the ion density along the shock front. The ion density oscillations are correlated with those of $B(x,y)$. 

The curves in Fig.~\ref{figure10}(c,~f) show that the average ion density downstream of the shock is about 2; this shock does not compress the plasma much because the shock speed in the downstream frame is not small compared to the relative speed between the downstream and upstream plasma. According to $P_{bx}$, deformations of $B(x,y)$ are strongest near the shock front. They extend upstream and downstream and decrease exponentially with distance from the maximum. Note that the curve has its maximum behind the front because the strong magnetic field does not fill out the simulation box for all values of $y$ at larger $x$. The peaks in the ion density have a much smaller spread along $x$ than their magnetic counterparts; charge density oscillations are shielded on Debye length scales and magnetic ones on skin depth scales. 

\begin{figure}
\includegraphics[width=\columnwidth]{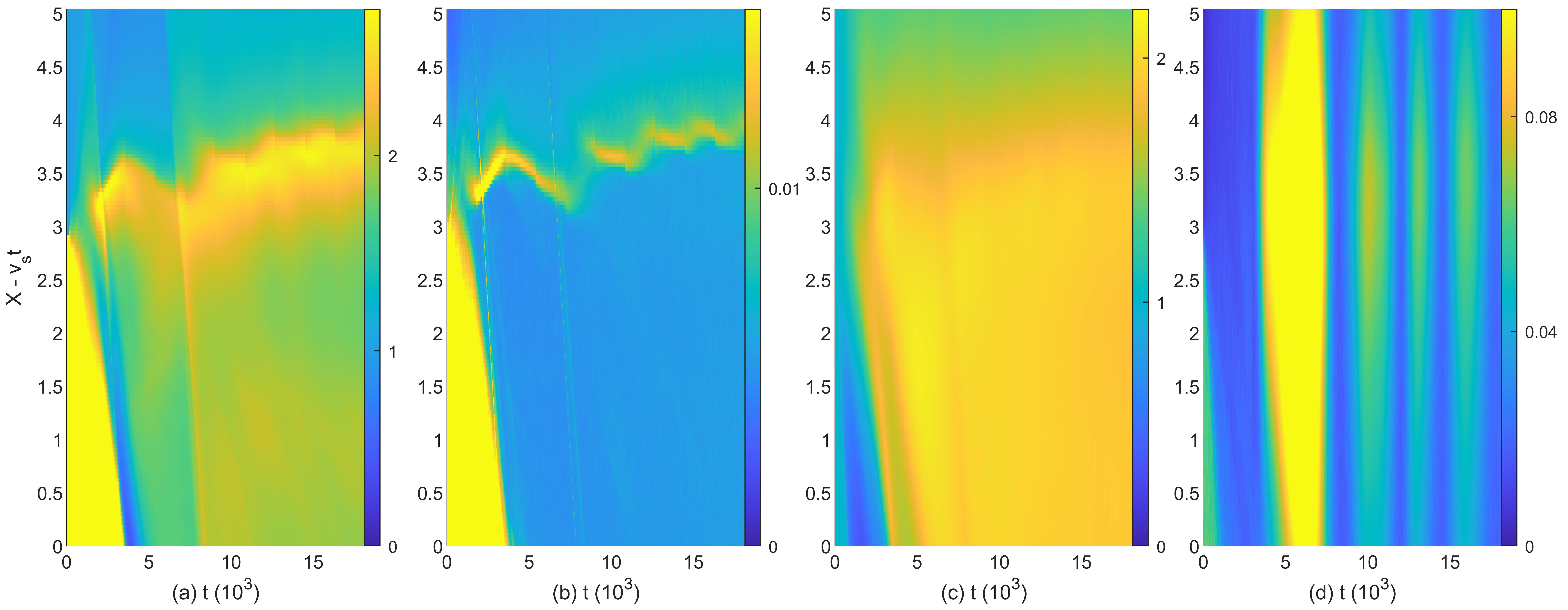}
\caption{Y-averaged ion density and electromagnetic fields in a reference frame $\tilde{x}=x-v_st$ that moves with $v_s = 1.6v_{fms}$. The shock is almost stationary in this reference frame. Panels~(a) and~(b) show the averaged ion density $N_i(\tilde{x},t)=L_y^{-1}\int n_i(\tilde{x},t) \, dy$ and the averaged amplitude modulus of the electric field in the simulation plane $\langle E^2 \rangle_y^{1/2}(\tilde{x},t) = (e/L_{y,3}m_ec\omega_{pe}){(\int_y (E_x(\tilde{x},t)^2 + E_y(\tilde{x},t)^2) \, dy)}^{1/2}$. Panels~(c) and~(d) show the average of the magnetic field modulus along $y$ ${\langle B_y^2 \rangle_y}^{1/2}(\tilde{x},t) = {(\int_y (B_y(\tilde{x},t)/B_0)^2 \, dy)}^{1/2}$ and along $x$ ${\langle B_x^2 \rangle_y}^{1/2}(\tilde{x},t) = {(\int_y (B_x(\tilde{x},t)/B_0)^2 \, dy)}^{1/2}$.}
\label{figure11}
\end{figure}

Figure~\ref{figure11} shows the evolution in time of the y-averaged density and field distributions in simulation~3. We have transformed these distributions into a reference frame that moves with the speed $v_s=1.6v_{fms}$ in the direction of the perturbed shock. Figure~\ref{figure11}(a) shows that the shock is located at $\tilde{x}=x-v_st\approx 3.5$. The position of its front oscillates in time. 
Its evolution is determined by a simultaneous oscillation of the shock front position, which varies with $y$ and $t$, and of the density along the front. A strong in-plane electric field marks the front of the shock in Fig.~\ref{figure11}(b) where the ion density gradient is largest. According to Fig.~\ref{figure11}(c), the magnetic field is amplified by the shock crossing to more than twice its upstream value. The contribution of $B_x$ to the magnetic pressure shown in  Fig.~\ref{figure11}(d) shows oscillations, which extend far into the upstream and downstream regions. The first and strongest maximum at early times is associated with the forced deformation of the shock as it propagated across the perturbation layer. After the shock left the perturbation layer, the shock performed free oscillations. They are coherent along the normal direction of the shock. Typical amplitudes of $B_x$ are weak compared to $B_y$ and we can thus interpret them as perturbations of an almost uniform guiding magnetic field that is aligned with $y$. 

\begin{figure}
\includegraphics[width=\columnwidth]{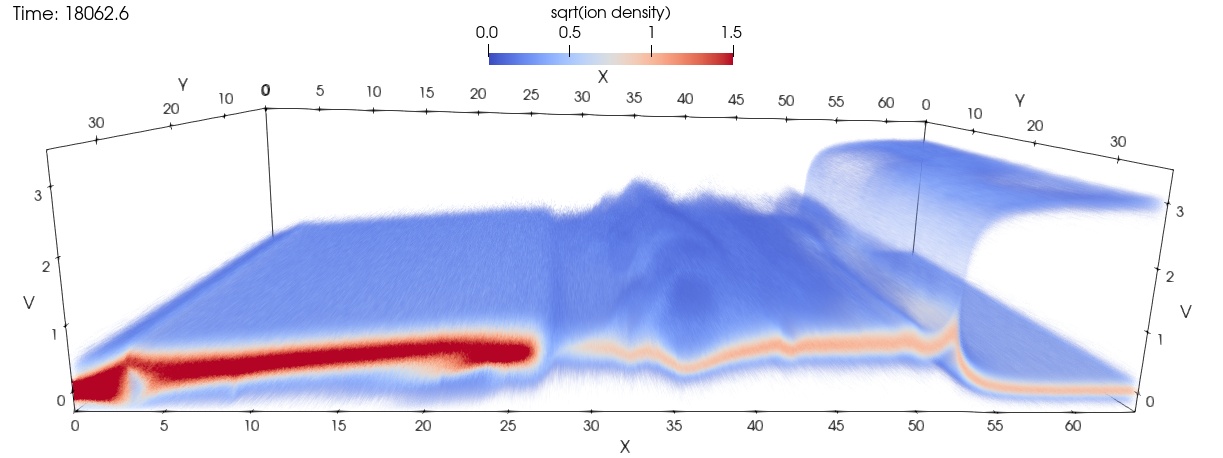}
\caption{The square root of the ion phase space density at the time $t=18000$ as a function of $x$, $y$, and $V=v_x/v_{fms}$. The density is normalized to the maximum phase space density of the ambient ions at the simulation's start and the color and opacity scales are linear.}
\label{figure12}
\end{figure}

The supplementary movie~5 animates the ion phase space density distribution in time for the interval $x>0$ and Fig.~\ref{figure12} shows its final frame. 
The shock reforms at the time $t\approx 1500$ and it remains stable after this time. It remains stationary in its comoving frame, which is typical for subcritical shocks. Some ions behind the shock are accelerated to high energies forming phase space vortices with an axis that is approximately aligned with the y-axis. One such vortex survives at $x\approx 36$ in Fig.~\ref{figure12}. A phase space vortex is sustained by the ambipolar electric field, which is associated with localized ion density depletion. The shock is located at $x\approx 53$ at this time and it accelerates a small fraction of the upstream ions to a speed $\approx 3v_{fms}$. The density of the shock-reflected ion beam is affected by the shock's density oscillation as can be seen from several patches with a reduced density. Since the upstream plasma outside of the perturbation layer has a uniform density, oscillations in the density of the shock-reflected ion beam must lead to a change in the number of ions that cross the shock and enter its downstream region. 

\section{Discussion}

We studied with PIC simulations in one and two spatial dimensions the stability of shocks in collisionless plasma. A pair of shocks was driven by the thermal pressure jump between a thin, dense, and planar plasma cloud in the center of the simulation box, and a surrounding dilute ambient plasma. One shock propagated through a spatially uniform ambient plasma. The other shock traversed a perturbation layer, in which we varied the number density of mobile ions after it fully formed. We explored the reaction of subcritical fast magnetosonic (FMS) shocks to the perturbation.

In the 1D simulations, a lower density of mobile ions upstream of the shock increased the speed of its downstream plasma and reduced the shock speed in the downstream frame of reference. For our initial conditions, the net effect was that the shock was slowed down in intervals with a lower number density of mobile ions. It regained its initial speed once it left the perturbation layer. The spatial lag between the fastest and the slowest shock was about 2 electron skin depths, which was comparable to the wavelength of the wave that mediated the shock. 

A variation in the density of the mobile ions in the perturbation layer along the shock front in the 2D simulations led to a spatial displacement of the shock front that was only a small fraction of the one we observed in the 1D PIC simulations. The amplitude of the spatial displacement was also proportional to the wavelength of the modulation; different parts of the shock front could thus not move independently. The crossing of the perturbation layer led to a modulation of the ion density of the shock front and to an expansion of the shocked magnetic field upstream. This expansion was more pronounced in regions with a low number density of mobile ions. Once a shock left the perturbation layer and entered the uniform plasma, the distributions of the ion density and position of the shock front, and the direction of the magnetic field changed.

We explored effects caused by the size of the simulation box along the shock front on the shock's evolution. The smallest box resolved a width of 12 electron skin depths, which is 6 times the wavelength of the wave that sustained the shock. The intermediate one doubled that width and the largest box tripled it. In the smallest simulation box, the shock perturbation damped out on a time scale less than an oscillation period of the mode that sustained the shock. In the intermediate simulation box, we observed a damped oscillation. Even weaker damping was observed in the largest box. 

The frequency, with which the shock perturbation oscillated, was a large fraction of the lower-hybrid frequency. This frequency matches that of the FMS waves that mediated the shock in a simulation with similar initial conditions~\cite{Dieckmann2017}. The ion density oscillations along the shock boundary and orthogonal to it constituted a standing wave, which is composed of modes that propagate in opposite directions and have a wavevector that is oblique to the shock normal. The limited thickness of the shock transition layer implies that the shock oscillation is not monochromatic.  Fast magnetosonic waves close to their resonance frequency are known as lower-hybrid waves and they are dispersive. This means that their phase velocity changes with the wavevector. The damping of the shock may thus be caused by different frequencies of the lower-hybrid modes that constitute the shock. The larger the simulation box along the shock boundary, the better the shock is resolved in wavenumber space and the lower the frequency mismatch of the shock modes becomes. This may explain why the damping of the shock perturbation became weaker with the increasing size of the simulation box.
 
All simulations have shown that the shocks remain stable during their traversal of the perturbation layer and in the uniform plasma. The boundary oscillations are thus damped. We could, however, not track the shock long enough in the largest 2D simulation to determine its damping rate and, more specifically, if the amplitude decreases in time $t$ with $t^{-3/2}$ as derived for some (magneto)hydrodynamic shocks. The boundary oscillations resulted in perturbations of the background magnetic field orthogonal to its direction. The background magnetic field was deformed in an interval that extended several electron skin depths upstream and downstream of the density overshoot of the shock in Figs.~\ref{figure10}(c, f) and in Fig.~\ref{figure11}(d). Such perturbations can trigger the growth of magnetowaves propagating in the Alfv\'en mode branch or in its high-frequency extension known as the Whistler wave branch~\cite{Graham2019}, which extends up to wave propagation angles that are almost perpendicular to the background magnetic field~\cite{Artemyev2016}. The damping of the shock boundary oscillations in the largest 2D simulation may be weak enough to lead to the growth of waves, which propagate along the background magnetic field. The box size and the simulation time we could resolve with our simulation were however not sufficiently large to resolve the  Alfv\'en waves, which were observed in hybrid simulations~\cite{Lowe2003}, in PIC simulations with different initial and boundary conditions~\cite{Yang2012, Clark2014, Umeda2017, Kobzar2021}, and by satellites~\cite{Johlander2016}. 

We could also not determine unambiguously the mechanism, that lets the shock oscillation saturate in the 2D perturbation layer. The observation that lower-hybrid waves can only mediate a quasi-perpendicular shock and that the perturbation oscillates at the lower-hybrid frequency suggests that the amplitude of the shock boundary oscillation is limited by the stability properties of lower-hybrid waves. Another saturation mechanism could be magnetic tension, which was small compared to the thermal pressure gradient force but comparable in magnitude to the magnetic pressure gradient force. We leave these studies to future work.

It would be interesting to investigate shock oscillations in the laboratory. A wide range of shock studies in collisionless plasma exist but to the best of our knowledge, none has examined oscillations of the shock boundary. One caveat to such studies is that the wavelength of these oscillations is long and their amplitude small. We selected the magnetic field amplitude 0.85 T, the electron temperature 1000 keV and density $10^{15}cm^{-3}$ and fully ionized nitrogen ions, because these values are realistic for experiments, in which a laser-generated blast shell expands into an ambient plasma. An example is the shock formation study in Ref.~\cite{Ahmed2013}. However, a wavelength of the perturbation in simulation~3 would amount to 5 mm, which exceeds by far the spatial scales that was resolved by that study.

\section*{Acknowledgements} The simulations were performed on resources provided by the Swedish National Infrastructure for Computing (SNIC) at the NSC and on the centers of the Grand Equipement National de Calcul Intensif (GENCI) under grant number A0090406960. MED acknowledges financial support from a visiting fellowship of the Centre de Recherche Astrophysique de Lyon. AB and FCC acknowledge support by grant PID2021-125550OB-I00 from the Spanish Ministerio de Economía y Competitividad.

\section*{Data availability statement} All data that support the findings of this study are included within the article (and any supplementary files).

\section*{Conflict of interest} The authors declare that they have no conflict of interest.

\end{document}